\documentclass[11pt]{article}
\usepackage{graphicx,epsfig,fullpage}

\newcommand {\E}{\mathbf E}

\renewcommand {\P}{\mathcal P}

\newtheorem{ex}{Example}[section]

\newcommand{\W}{\mathcal{W}}
\newcommand{\eat}[1]{}
\renewcommand{\paragraph}[1]{\medskip \noindent {\bf{#1}}}

\newcommand{\qed}{\mbox{}\hspace*{\fill}\nolinebreak\mbox{$\rule{0.6em}{0.6em}$}}
\newtheorem{theorem}{Theorem}[section]
\newtheorem{lemma}[theorem]{Lemma}
\newtheorem{claim}[theorem]{Claim}

\newenvironment{proof}{{\bf Proof:}}{$\qed$\par}

\newcommand{\B}{\mathcal{B}}

\usepackage{hyperref}
\hypersetup{
bookmarksnumbered
}

\begin{document}
\title{Hybrid Keyword Search Auctions\footnote{To appear in the proceedings of
    ACM WWW '09}}
\author{
  Ashish Goel\thanks{Departments of Managment Science and Engineering and (by
    courtesy) Computer Science, Stanford University. Research supported in part by NSF ITR grant 0428868, by gifts from Google, Microsoft, and
    Cisco, and by the Stanford-KAUST alliance. Email: ashishg@stanford.edu .}\\
  Stanford University
  \and
  Kamesh Munagala\thanks{Department of Computer Science, Duke University.
    Research supported by NSF via a CAREER award and grant CNS-0540347.
    Email: kamesh@cs.duke.edu .}\\Duke University
}
\date{\today}


\maketitle
\pdfbookmark[1]{Abstract}{MyAbstract}
\begin{abstract}
  Search auctions have become a dominant source of revenue
  generation on the Internet. Such auctions have typically
  used per-click bidding and pricing. We propose the use of
  hybrid auctions where an advertiser can make a
  per-impression as well as a per-click bid, and the
  auctioneer then chooses one of the two as the pricing
  mechanism. We assume that the advertiser and the
  auctioneer both have separate beliefs (called priors) on
  the click-probability of an advertisement. We first prove
  that the hybrid auction is truthful, assuming that the
  advertisers are risk-neutral. We then show that this
  auction is superior to the existing per-click auction in
  multiple ways:
  \begin{enumerate}
  \item We show that risk-seeking advertisers will choose
    only a per-impression bid whereas risk-averse
    advertisers will choose only a per-click bid, and argue
    that both kind of advertisers arise naturally. Hence,
    the ability to bid in a hybrid fashion is important to
    account for the risk characteristics of the advertisers.
  \item For obscure keywords, the auctioneer is unlikely to
    have a very sharp prior on the click-probabilities. In
    such situations, we show that having the extra
    information from the advertisers in the form of a
    per-impression bid can result in significantly higher
    revenue.
  \item An advertiser who believes that its
    click-probability is much higher than the auctioneer's
    estimate can use per-impression bids to correct the
    auctioneer's prior {\em without incurring any extra
      cost.}
  \item The hybrid auction can allow the advertiser and
    auctioneer to implement complex dynamic programming
    strategies to deal with the uncertainty in the
    click-probability using the same basic auction. The
    per-click and per-impression bidding schemes can only be
    used to implement two extreme cases of these strategies.
  \end{enumerate}
  As Internet commerce matures, we need more sophisticated
  pricing models to exploit all the information held by each
  of the participants. We believe that hybrid auctions could
  be an important step in this direction. The hybrid auction
  easily extends to multiple slots, and is also applicable
  to scenarios where the hybrid bidding is per-impression
  and per-action (i.e.  CPM and CPA), or per-click and
  per-action (i.e. CPC and CPA).
\end{abstract}





\section{Introduction}
\label{sec:intro}

While search engines had a transformational effect on Internet use and indeed,
on human interaction, it was only with the advent of keyword auctions that
these search engines became commercially viable. Most of the major search
engines display advertisements along with search results; the revenue from
these advertisements drives much of the innovation that occurs in search in
particular, and Internet applications in general. Cost-per-click (CPC) auctions
have evolved to be the dominant means by which such advertisements are
sold~\cite{v:google06}. An advertiser places a bid on a specific keyword or
keyword group. The auctioneer (i.e. the search engine) maintains an estimate
of the click-through probability (CTR) of each advertiser for each
keyword. When a user searches for a keyword, the first advertising spot is
sold to the advertiser which has the highest product of the bid and the CTR;
in the event that this advertisement is clicked upon by the user, this
advertiser is charged the minimum bid it would have to make to retain its
position. The same process is repeated for the next slot, and so on. A full
description of the tremendous amount of work related to keyword auctions is
beyond the scope of this paper; the reader is referred to the excellent survey
by Lahaie et al~\cite{lpsv:auctions07}. Two other alternatives to CPC auctions are
widely used:
\begin{enumerate}
\item CPM, or Cost Per (thousand) Impressions: The publisher charges the
  advertiser for every instance of an advertisement shown to a user,
  regardless of the click.  This is widely used to sell banner advertisements.
\item CPA, or Cost Per Action (also known as Cost Per Acquisition): The
  publisher charges the advertiser when an actual sale happens. This is widely
  used in associate programs such as the ones run by Amazon, and by lead
  generation intermediaries.
\end{enumerate}

The three models are equivalent when precise estimates of the
click-through-probability and click-to-sale-conversion probability are
known. In the absence of such estimates, CPC has emerged as a good way of
informally dividing the risk between the auctioneer and the advertiser: the
auctioneer is vulnerable only to uncertainty in its own estimates of CTR,
whereas the advertiser is vulnerable only to uncertainty in its own estimates
of the click-to-sale-conversion probability, assuming its advertisement gets
displayed in a favorable spot. A great deal of effort has gone into obtaining
good predictions of the CTR. The problem is made harder by the fact that many
keywords are searched for only a few times, and typical CTRs are
low. Advertisers often want to also bid by customer demographics, which
further exacerbates the sparsity of the data. Hence, there has to be a great
reliance on predictive models of user behavior and new ads (e.g. see
~\cite{rdr:ctr07}). Arguably, another approach would to devise pricing models
that explicitly take the uncertainty of the CTR estimates into account, and
allow advertisers and auctioneers to jointly optimize over this
uncertainty. In general, we believe that as Internet commerce matures, we need
not just better estimation and learning algorithms but also more sophisticated
pricing models to exploit all the information held by each of the
participants.

In this paper, we propose the use of hybrid auctions for search keywords where
an advertiser can make a per-impression as well as a per-click bid, and the
auctioneer then chooses one of the two as the pricing mechanism. Informally,
the per-impression bids can be thought of as an additional signal which
indicates the advertiser's belief of the CTR. This signal may be quite
valuable when the keyword is obscure, when the advertiser is aggregating data
from multiple publishers or has a good predictive model based on domain
knowledge, and when the advertiser is willing to pay a higher amount in order
to perform internal experiments/keyword-selection. We assume that the
advertiser and the auctioneer both have separate beliefs (called priors) on
the click-probability of an advertisement.

\subsection{Our results}

We describe the hybrid auction in section~\ref{sec:hybrid}, where we also
outline the strategic model (that of discounted rewards) used by the
auctioneer and the advertiser. We introduce the multi-armed bandit problem as
it occurs naturally in this context. Our results, which we have already
described at a high level in the abstract, are split into two parts.

\paragraph{Myopic advertisers.}
We first study (section~\ref{sec:myopic}) the case of myopic advertisers,
which only optimize the expected profit at the current step. When these
advertisers are risk-neutral, we show that truth-telling is a strongly
dominant strategy: the advertiser bids the expected profit from an impression
as a per-impression bid, and the value it expects from a click as the
per-click bid, regardless of the auctioneer's prior or optimization
strategy. Further, if the advertiser is certain about its CTR, and if this CTR
is drawn from the auctioneer's prior which follows the natural Beta
distribution (defined later), then the worst case loss in revenue of the
auctioneer over pure per-click bidding is at most $1/e \approx 37\%$. In
contrast, the revenue-loss for the auctioneer when he uses the per-click
scheme as opposed to the hybrid auction {\em can approach 100\% for a fairly
  natural scenario, one that corresponds to obscure keywords}. We finally
consider risk taking behavior of the advertisers when they are not certain
about their CTR. We show that per-click bidding is dominant when the
advertisers are risk-averse, but per-impression bidding is desirable when they
are risk-seeking. Thus, the hybrid auction
\begin{enumerate}
\item Naturally extends the truthfulness of the single-slot per-click bidding
  auctions currently in use, for the case of myopic, risk-neutral advertisers
  (which is the situation under which the properties of the per-click auction
  are typically analyzed~\cite{v:google06,edel-05,agm:auction06,msvv-05}).
\item Results in bounded possible revenue loss but unbounded possible revenue
  gain for the auctioneer in the natural setting of risk-neutral, myopic
  advertisers, and where the auctioneer uses the Gittins index. The revenue
  gain occurs in the common setting of obscure keywords.
\item Naturally takes the risk posture of the advertiser into account, which
  neither per-click nor per-impression bidding could have done on its own
  (both risk averse and risk seeking advertisers occur naturally).
\end{enumerate}
The result bounding the possible revenue loss of the auctioneer under the
hybrid auction is for an arbitrary discount factor used by the auctioneer; the
results about the possible revenue gain and the risk posture assume a myopic
auctioneer. We believe these are the most appropriate assumptions, since we
want to provide bounds on the revenue loss using hybrid auctions under the
most general scenario, and want to illustrate the benefits of using the hybrid
auction under natural, non-pathological scenarios.

\paragraph{Semi-myopic advertisers.}
In section~\ref{sec:semi-myopic}, we remove the assumption that the
advertisers are only optimizing some function of the profit at the current
step. We generalize to the case where the advertisers optimize revenue over a
time-horizon. We develop a tractable model for the advertisers, and show a
simple dominant strategy for the advertisers, based on what we call the {\em
  bidding index}. Though this strategy does not have a closed form in general,
we show that in many natural cases (detailed later), it reduces to a natural
pure per-click or pure per-impression strategy that is socially optimal. Thus,
our hybrid auctions are flexible enough to allow the auctioneer and the
advertiser to implement complex dynamic programming strategies
collaboratively, under a wide range of scenarios. Neither per-impression nor
per-click bidding can exhaustively mimic the bidding index in these natural
scenarios.

Finally, we show a simple bidding strategy for a certain (i.e. well-informed)
advertiser to make the auctioneer's prior converge to the true CTR, while
incurring no extra cost for the advertiser; per-click bidding would have
resulted in the advertiser incurring a large cost. This is our final argument
in support of hybrid auctions, and may be the most convincing from an
advertiser's point of view.

We explain throughout the paper why the scenarios we consider are not
arbitrarily chosen, but are quite natural (indeed, we believe the most natural
ones) to analyze. In the process, we obtain many interesting properties of the
hybrid auction, which are described in the technical sections once we have the
benefit of additional notation.

\paragraph{Multiple Slots.} The main focus of the paper is analyzing the
properties of the {\sc hybrid} scheme on a single ad slot. However, the
auction itself can be easily generalized to multiple slots in two different
ways; before describing these, we need to note that the hybrid auction assigns
an ``effective bid'' to each advertiser based on the per-impression bid, the
per-click bid, and the expected CTR or quality measure. The first
generalization is akin to the widely used generalized second price
auction~\cite{edel-05,v:google06} (also referred to as the ``next-price''
auction~\cite{agm:auction06}) for CPC-only bidding: the advertisers are ranked
in decreasing order of effective bids, and the ``effective charge'' made to
each advertiser is the effective bid of the next advertiser. We do not discuss
this variant in the rest of this paper, since the computation methodology is
no different from single slot auctions. Note that we can not expect this
multi-slot generalization of the hybrid auction to be truthful because even
the CPC-only next-price auction is not
truthful~\cite{v:google06,edel-05,agm:auction06}. However, given the immense
popularity of the next-price auction, we believe that this generalization of
the hybrid auction is the most likely to be used in real-life settings.

The second generalization mirrors VCG~\cite{edel-05} (or equivalently, a
laddered CPC auction~\cite{agm:auction06}).  This generalization assumes that
the CTR is multiplicatively separable into a position dependent term and an
advertiser dependent term, and under this assumption, guarantees truthfulness
(on both the per-impression and per-click bids) for myopic, risk-neutral
advertisers.  Details of this auction are in section~\ref{sec:multi}. The
proof follows by extending the proof of theorem~\ref{thm:truth} exactly along
the lines of~\cite{agm:auction06} and is omitted.

The hybrid auction is also applicable to scenarios where the hybrid bidding is
per-impression and per-action (i.e.  CPM and CPA), or per-click and per-action
(i.e. CPC and CPA).

\section{The Hybrid Auction Scheme}
\label{sec:hybrid}
As mentioned before, we will assume that there is a single slot that is being
auctioned.  There are $n$ advertisers interested in a single keyword. When an
advertiser $j$ arrives at time $t=0$, it submits a bid $(m_{jt},c_{jt})$ to
the advertiser at time-slot $t \ge 1$. The interpretation of this bid is that
the advertiser is willing to pay at most $m_{jt}$ per impression or at most
$c_{jt}$ per click. These values are possibly conditioned on the outcomes at
the previous time slots. The auctioneer chooses a publicly known
value\footnote{It is conceivable that an auctioneer may strategically decide
  to not reveal its true prior; this would be an interesting direction to
  consider in future work.}  $q_{jt}$ which we term the {\em auctioneer
  index}, which can possibly depend on the outcomes for this advertiser at the
previous time slots when its advertisement was shown, but is independent of
all the bids.

The {\sc Hybrid} auction scheme mimics VCG on the quantity $R_{jt} =
\max(m_{jt}, c_{jt} q_{jt})$. We will call $R_{jt}$ the effective bid of user
$j$ at time $t$. Let $j^*$ denote the advertiser with highest $R_j$ value, and
$R_{-j^*}$ denote the second highest $R_j$. There are two cases.  First,
suppose $m_{j^*} > c_{j^*} q_{j^*}$, then $j^*$ gets the slot at
per-impression price $R_{-j^*}$. In the other case, $j^*$ gets the slot at
per-click price $\frac{R_{-j^*}}{q_{j^*}}$. It is clear that the {\sc Hybrid}
scheme is feasible, since the per-impression price charged to $j$ is at most
$m_j$, and the per-click price is at most $c_j$. The auction generalizes in a
natural way to multiple slots, but we will focus on the single slot case in
this paper. If the auctioneer chooses $q_{j,t}$ to be an estimate of the
click-through-rate (CTR) and the advertiser submits only a per-click bid, then
this reduces to the traditional next-price auction currently in use.

In order to analyze properties of the {\sc Hybrid} auction, we need to make
modeling assumptions about the advertiser, about the auctioneer index $q$, and
about time durations.

\subsection{Time Horizon and the Discount Factor}
\label{sec:modeldiscount}
To model the time scale over which the auction is run, we assume there is a
global discount factor $\gamma$. Informally, this corresponds to the present
value of revenue/profit/cost that will be realized in the next
step, and is an essential parameter in determining the correct tradeoff
between maximizing present expected reward (exploitation) vs. obtaining more
information with a view towards improving future rewards (exploration).
The expected revenue at time step $t$ gets multiplied by a factor of
$\gamma^t$.
Note that $\gamma = 0$ corresponds to optimizing
for the current step (the {\em myopic} case). In the discussion below, we
assume the auctioneer and advertiser behave strategically in optimizing their
own revenues, and can compute parameters and bids based on their own discount
factors which could be different from the global discount factor $\gamma$ used
for discussing social optimality.

\renewcommand{\P}{\mathcal{P}}
\newcommand{\Q}{\mathcal{Q}}

\subsection{Auctioneer Model and the Gittins Index}
\label{sec:modelauctioneer}
For the purpose of designing an auctioneer index $q_{jt}$, we assume the
auctioneer starts with a prior distribution $\Q_j$ on the CTR of advertiser
$j$. We assume further that he announces this publicly, so that the advertiser
is aware of this distribution. Therefore, the a priori expected value of the
CTR of advertiser $i$ from the point of view of the auctioneer is $ \E[\Q_j]$.
Suppose at some time instant $t$, $T_{jt}$ impressions have been offered to
advertiser $j$, and $n_{jt}$ clicks have been observed. The natural posterior
distribution $\Q_{jt}$ for the advertiser is given by:
$$\Pr[\Q_{jt} = x] \propto x^{n_{jt}} (1-x)^{T_{jt} - n_{jt}} \cdot
\Pr[\Q_j = x];$$ this corresponds to Bayesian updates, and when initialized
with the uniform continuous prior, corresponds to the natural Beta
distribution, defined later. The auctioneer chooses a function $f(\Q_{jt})$
that maps a posterior distribution $\Q_{jt}$ to a $q$ value.

The function $f$ is chosen based on the revenue guarantees the auctioneer
desires. We will use the following idealized scenario to {\em illustrate} a
concrete choice of $f$; our results apply broadly and are not limited to the
scenario we describe. Suppose the auctioneer wishes to optimize over a time
horizon given by discount factor $\gamma_a$. Then, if the auctioneer were to
ignore the per impression bids, and uses a first price auction on $c_j q_{jt}$
assuming that $c_j$ is the true per-click valuation, then his choice of
$q_{jt}$ should maximize his expected discounted revenue. Let $v_j$ denote the
true per-click valuation of the advertiser. By assumption, $v_j = c_j$. At
time $t$, the auctioneer offers the slot to the bidder $j^*$ with highest $v_j
q_{jt}$ at per-click price $v_{j^*}$, earning $v_{j^*} \E[\Q_{jt}]$ in
expectation. It is well-known that the discounted reward of this scheme is
maximized when $q_{jt}$ is set to the {\em Gittins index} (described next) of
$\Q_{jt}$ with discount factor $\gamma_a$, and hence setting $f(\Q_{jt}) = $
the Gittins index of $\Q_{j,t}$ would be a natural choice.

The {\em Gittins index}~\cite{g:gittins79,w:gittins92} of a distribution $\Q$
for discount factor $\gamma_a$ is defined as follows: Consider a coin with
this prior distribution on probability of heads $\Q$, and that yields reward
$1$ on heads. The Gittins index is $(1-\gamma) M$, where $M$ is the largest
number satisfying the following condition: Some optimal discounted reward
tossing policy that is allowed to retire at any time point and collect a
retirement reward of $M$ will toss the coin at least once. It is well-known
that the Gittins index is at least the mean $\E[\Q]$ of the prior, and for a
given mean, the Gittins index increases with the variance of the prior, taking
the lowest value equal to the mean only when the prior has zero
variance. Further, the Gittins index also increases with the discount factor
$\gamma_a$, being equal to the mean when $\gamma_a = 0$.

An equivalent definition will be useful: Consider a coin with the prior
distribution on probability of heads $\Q_{j,t}$ that yields reward $1$ on
heads. Suppose the coin is charged $G$ amount whenever it is tossed, but is
allowed to retire anytime. The Gittins index is the largest $G$ for which the
expected discounted difference between the reward from tossing minus the
amount charged in the optimal tossing policy is non-negative\footnote{The
  Gittins index is usually defined as $M$ (i.e., $G/(1-\gamma)$) but the
  alternate definition $(1-\gamma)M$ (i.e., $G$) is more convenient for this
  paper.}.

Typically, the distribution $\Q$ is set to be the conjugate of the Bernoulli
distribution, called the Beta distribution~\cite{g:gittins79}. The
distribution Beta$(\alpha, \beta)$ corresponds to starting with a uniform
distribution over the CTR and observing $\alpha-1$ clicks in $\alpha+\beta -
2$ impressions. Therefore, if the initial prior is Beta$(\alpha,\beta)$, and
$n$ clicks are then observed in $T$ impressions, the posterior distribution is
Beta$(\alpha+n,\beta+T-n)$. Beta$(1,1)$ corresponds to the uniform
distribution. Beta distributions are widely used mainly because they are easy
to update. However, unless otherwise stated, our results will not depend on
the distribution $\Q$ being a Beta distribution.

\subsection{Advertiser Model}
\label{sec:modeladvertiser}
The value bid by the advertisers will depend on their optimization criteria.
The true per-click value of advertiser $j$ is $v_j$. The advertiser $j$
maintains a time-indexed distribution $\P_{jt}$ over the possible values of
the actual CTR $p$ that is updated whenever he receives an impression. We
assume advertiser $j$'s prior is updated based on the observed clicks in a
fashion similar to the auctioneer's prior, but again, this is not essential to
our results except where specifically mentioned.

The advertiser's bid will depend on its optimization criteria. In the
next section, we consider the case where the advertisers only optimize
their revenue at the current step, and could possibly take risk. In
later sections, we consider the case where the advertisers attempt to
optimize long-term revenue by bidding strategically over time. 

In each case, the advertiser could be well-informed (or certain) about its
CTR, so that $\P_{jt}$ is a point distribution, or uninformed about its CTR,
so that it trusts the auctioneer's prior, {\em i.e.}, $\P_{jt} = \Q_{jt}$, or
somewhere in between. Depending on the optimization criterion of the
advertiser, these cases lead to different revenue properties for the
auctioneer and advertiser, and show the advantages of the {\sc Hybrid} scheme
over pure per-click bidding as well as over per-impression bidding.

\section{Myopic Advertisers}
\label{sec:myopic}
In this section, we analyze single time-step properties of the
auction. Specifically, we assume that the advertisers are myopic,
meaning that they optimize some function of the revenue at the current
time step. 

Since we consider myopic properties, we drop the subscript $t$
denoting the time step from this section. The auctioneer's prior is
therefore $\Q_j$, and the advertiser's prior is $\P_j$. Let $p_j =
\E[\P_j]$. 

We first show that when the advertisers are risk-neutral, then bidding $( v_j
p_j, v_j )$ is the dominant strategy, {\em independent of the auctioneer's
  prior or the choice of $f$}. Further, if the advertiser is certain about its
CTR, and if this CTR is drawn from the auctioneer's prior which follows a Beta
distribution, then the worst case loss in revenue of the auctioneer over pure
per-click bidding is at most $1/e \approx 37\%$. In contrast, the revenue-loss
for the auctioneer when he uses the per-click scheme as opposed to the hybrid
auction {\em can approach 100\% for a fairly natural scenario, one that
  corresponds to obscure keywords}. We finally consider risk taking behavior
of the advertisers when they are not certain about their CTR. We show that
per-click bidding is dominant when the advertisers are risk-averse, but
per-impression bidding is desirable when they are risk-seeking. Thus, the
hybrid auction naturally extends the truthfulness of the single-slot per-click
bidding auctions currently in use, results in bounded possible revenue loss
but unbounded possible revenue gain, and naturally takes the risk posture of
the advertiser into account; the precise qualitative conclusions are
detailed in the introduction and the formal statements are proved below.

The result bounding the possible revenue loss of the auctioneer under the
hybrid auction is for an arbitrary discount factor used by the auctioneer; the
results about the possible revenue gain and the risk posture assume a myopic
auctioneer. We believe these are the most appropriate assumptions, since we
want to provide bounds on the revenue loss using hybrid auctions under the
most general scenario, and want to illustrate the benefits of using the hybrid
auction under natural, non-pathological scenarios.
\subsection{Truthfulness}
\label{sec:truthful}
We first show that the dominant strategy involves truthfully revealing
the expected CTR, $p_j$.  Recall that the advertiser bids $(m_j,
c_j)$. Further, the auctioneer computes an index $q_j$ based on the
distribution $\Q_j$, and does VCG on the quantity $R_j = \max(m_j, c_j
q_j)$.

\begin{theorem}
\label{thm:truth}
If $p_j = \E[\P_j]$ and the advertiser is myopic and risk-neutral,
then regardless of the choice of $q_{j}$, the (strongly) dominant
strategy is to bid $( v_j p_j, v_j )$.
\end{theorem}
\begin{proof}
  First, consider the case where $q_j \le p_j$. Suppose the advertiser bids
  $(m_j, c_j)$ and wins the auction. Then, the expected profit of this
  advertiser is at most $p_jv_j - \min \{R_{-j*}, R_{-j*}\cdot(p_j/q_j)\}$
  which is at most $p_jv_j - R_{-j*}$. Thus, the maximum profit of the
  advertiser can be at most $\max\{0,p_jv_j - R_{-j*}\}$ which is obtained by
  bidding $(p_jv_j, v_j)$.

  Next, consider the case where $q_j > p_j$. Suppose the advertiser bids
  $(m_j, c_j)$ and wins the auction. Then, the expected profit of this
  advertiser is at most $p_jv_j - \min \{R_{-j*}, R_{-j*}\cdot(p_j/q_j)\}$
  which is at most $p_jv_j - R_{-j*}\cdot(p_j/q_j)$. Thus, the maximum
  profit of the advertiser can be at most $\max\{0,p_jv_j -
  R_{-j*}\cdot(p_j/q_j)\}$ which is again obtained by bidding $(p_jv_j,
  v_j)$.

  Thus, it is never suboptimal to bid truthfully. Let $R^*$ denote the second
  highest value of $\max(m_j,c_j q_j)$.  In order to show that $(p_jv_j, v_j)$
  is a (strongly) dominant strategy, we need to show that for any other
  bid-pair $(m_j, c_j)$, there exist values of $q_j, R^*$ such that the profit
  obtained by bidding $(m_j, c_j)$ is strictly less than that obtained by
  truthful bids. Suppose $\epsilon$ is an arbitrary small but positive
  number. First consider the scenario where $q_j = p_j$, i.e., the auctioneer
  has a perfect prior. In this scenario, bidding $(m_j, c_j)$ with either $m_j
  > p_j v_j + \epsilon$ or $c_j > v_j + \epsilon/p_j$ results in a negative
  profit when $p_jv_j < R^* < p_jv_j + \epsilon$; truthful bidding would have
  resulted in zero profit.  Further, if $m_j < p_jv_j - \epsilon$ and $c_j <
  v_j - \epsilon/p_j$, then the advertiser obtains zero profit in the case
  where $p_jv_j > R^* > p_jv_j - \epsilon$; truthful bidding would have
  obtained positive profit.

  This leaves the cases where $m_j = p_jv_j, c_j < v_j/(1+\epsilon)$ or where
  $m_j < p_jv_j/(1+\epsilon), c_j = v_j$. In the former case, the advertiser
  obtains zero profit in the situation where $q_j > p_j > q_j/(1+\epsilon)$
  and $R^* = p_jv_j$; truthful bidding would have obtained positive profit. In
  the latter case, the advertiser obtains zero profit when $q_j <
  p_j/(1+\epsilon)$ and $R^* = p_jv_j/(1+\epsilon)$; truthful bidding would
  have obtained positive profit.
\end{proof}

\subsection{Well-Informed Advertisers: Loss in Auctioneer's Revenue}  
\label{sec:lossauctioneer}
We now consider the case where the advertisers are certain about their CTR
$p_j$ and risk-neutral; by the results of the previous section, we will assume
that they bid truthfully. More formally, we assume the prior $\P_{j}$ is the
point distribution at $p_j$. We suppose that the $p_j$ are drawn from the
auctioneer's prior that is of the form $\Q_j=$ Beta$(\alpha_j,\beta_j)$.  We
now show that for $q_j$ being the Gittins index of Beta$(\alpha_j,\beta_j)$ for
{\em any} discount factor $\gamma_a$, the expected revenue of the auctioneer
at the current step is at least $1-1/e$ times the revenue had he ignored the
per-impression bids. 
\begin{theorem}
\label{thm:63}
In the above mentioned scenario, the expected revenue of the
auctioneer at the current step is at least $1-1/e \approx 63\%$ of the
corresponding auction that ignores the per-impression bid.
\end{theorem}
\begin{proof}
  Let $q_j$ denote the Gittins index of $\Q_j=$Beta$(\alpha_j,\beta_j)$. Let
  advertiser $1$ have the highest $v_j q_j$, and advertiser $2$ the next
  highest.  Let $R^* = v_2
  q_2$. If the per-impression bids are ignored, advertiser $1$ gets the
  impression at a per-click price of $v_2 q_2/q_1$, so that the expected
  revenue is $R^* \frac{\E[\Q_1]}{q_1}$.

  In the {\sc Hybrid} scheme, $v_1q_1$ and $v_2q_2$ are both at least as large
  as $R^*$. Hence, if the auctioneer makes a per-impression charge, then this
  charge must be at least $R^*$ per impression. If the advertiser makes a
  per-click charge (which must be to advertiser $1$), then the expected revenue is at least
  $R^*\Q_1/q_1$. Hence the expected revenue of the {\sc Hybrid} scheme is at
  least $R^* \E\left[\min\left(1,\frac{\Q_1}{q_1}\right)\right]$ and the ratio of
  the revenue of the {\sc Hybrid} scheme to the per-click scheme is at least
  $\frac{\E[\min(q_1,\Q_1)]}{\E[\Q_1]}$.

 For $p$ drawn from
  the distribution $\Q_1$, we now need to show that
  $\frac{\E[\min(q_1,\Q_1)]}{\E[\Q_1]} \ge 1-1/e$. To show this, observe that
  for a fixed $\Q_1$, this ratio is smallest when $q_1$ is as small as
  possible. This implies we should choose $q_1 = \E[\Q_1] =
  \frac{\alpha}{\alpha+\beta}$, which corresponds to a discount factor of
  0. Denote $\mu = \E[\Q_1]$. Then, the goal is to minimize the ratio
  $\frac{1}{\mu}\E[\min(\mu,\Q_1)]$ as a function of
  $\alpha,\beta$. Lemma~\ref{lem:maj} shows that this ratio is $1-1/e$,
  completing the proof.
\end{proof}

\begin{lemma}
\label{lem:maj}
  If $w$ is drawn from the Beta distribution with parameters
  $\alpha,\beta \ge 1$, and $\mu = \alpha/(\alpha+\beta)$ is the mean
  of $w$, then $\E[\min(\mu,w)] \ge \mu(1-1/e)$.
\end{lemma}
\begin{proof}
  We will allow $\alpha,\beta$ to take on fractional values as long as they
  are both at least 1. Suppose $\alpha,\beta$ are both strictly bigger than 1.
  Let $z$ denote the random variable drawn from the Beta distribution with
  parameters $\alpha' = \alpha - \mu\theta$, $\beta' = \beta - (1-\mu)\theta$,
  where $\theta > 0$ is chosen such that $\alpha',\beta' \ge 1$ and at least
  one of $\alpha',\beta'$ is exactly 1. The mean of $z$ is ${\alpha -
    \mu\theta\over \alpha + \beta - \theta} = \mu$, which is the same as the
  mean of $w$.

  Let $f_w,f_z$ denote the probability density functions of $w, z$
  respectively, and let $F_w(x)$ (resp. $F_z(x)$) denote $\Pr[w \ge
  x]$ (resp. $\Pr[z \ge x]$). Consider the ratio $r(x) = f_w(x)/f_z(x)
  = \phi x^{\mu\theta}(1-x)^{(1-\mu)\theta}$, where $\phi$ is the
  ratio of the corresponding normalizing terms and hence does not
  depend on $x$.

  Observe that $r(x)$ is uni-modal (since the derivative of $r$ is 0 exactly
  once in the interval $[0,1]$); and that $r(x) \rightarrow 0$ as
  $x\rightarrow 0^+$ and as $x \rightarrow 1^-$.  Since both $F_w(x)$ and
  $F_z(x)$ are monotonically decreasing curves connecting $(0,1)$ and $(1,0)$,
  the above properties of $r(x) = \frac{F'_w(x)}{F'_z(x)}$ easily imply that
  for some $s \in (0,1)$, over the interval $x \in [0,s]$, $F_w(x) \ge
  F_z(x)$, and over $x \in [s,1]$, $F_w(x) \le F_z(x)$.  This combined with
  the fact that $\E[w] = \E[z] = \mu$ implies $w$ Lorenz-dominates $z$, so
  that for all concave functions $g$, we have $\E[g(w)] \ge
  \E[g(z)]$~\cite{mo:major79}.

Since $g(w) = \min(w,y)$ is concave in $w$ for fixed $y$, we have
$\E[\min(w,\mu)] \ge \E[\min(z,\mu)]$. Therefore, it is sufficient to
analyze $\E[\min(\mu,z)]/\mu$, {\em i.e.} the case where either
$\alpha$ or $\beta$ is exactly 1, and the other is at least 1.  It is
easy to explicitly verify both these cases, and show that the worst
case is when $\alpha = 1$ and $\beta \rightarrow \infty$ when
$\E[\min(\mu,z)] = (1-1/e) \mu$.
\end{proof}

\newcommand{\expect}{\E}

\paragraph{A Typical Case.} Though the {\sc Hybrid} scheme is not revenue
dominant over the pure per-click scheme in pathological cases, the key
advantage is in the following typical situation. There are $n$ advertisers
whose CTRs $p_1 \ge p_2 \ge \cdots \ge p_n$ are drawn from a common prior
$\Q=$ Beta$\left(1,K\right)$, whose mean is $\mu =
  O\left(\frac{1}{K}\right)$. Assume further that $n = 4^K$ or $K = \frac{\log
    n}{2}$. We have:
\begin{eqnarray*}
\Pr[p_2 \ge 1/2]  & \ge & \left(1- \left(1 - \frac{2}{\log n}
    \left(\frac{1}{2}\right)^{\frac{\log n}{2}}\right)^{\frac{n}{2}}\right)^2 \\
& = &\left(1- \left(1- \frac{2}{\sqrt{n}{\log n}}\right)^{\frac{n}{2}}\right)^2 \\
& = &1-o(1)
\end{eqnarray*} 
Recall that the advertisers are aware of their CTR, but the
auctioneer is only aware of the prior. Suppose the per-click value for all the
advertisers is $v$, and these are truthfully revealed.  In the per-click
scheme, the auctioneer sells the impression to an arbitrary advertiser at
per-click price $v$, and in expectation earns $\mu v$. If the auctioneer is
myopic ($\gamma_a = 0$), then $q =\E[\Q] < p_2$ w.h.p, and the {\sc Hybrid}
scheme charges per-impression. Here, the auctioneer sells to advertiser $1$ at
per-impression price $v p_2$. From the above, $\E[p_2] = \Omega(1)$, so that
$\E[p_2]/\mu = \Omega(\log n)$. Therefore, for $n$ advertisers with diffuse
priors of the form Beta$\left(1,\frac{1}{\log n}\right)$, the auctioneer gains
a factor $\Omega(\log n)$ in revenue. This is particular relevant for obscure
keywords, where the auctioneer will have very diffuse priors.

\subsection{Uninformed Advertisers and Risk}
\label{sec:riskadvertiser}
So far, we have assumed that the advertiser is risk neutral and certain about
the CTR, so that he is optimizing his expected profit. We now suppose that the
advertiser is uncertain and trying to maximize a utility function $U$ on his
profit. The function $U(x)$ is monotone with monotone derivative, and $U(0) =
0$. If $U$ is convex, the advertiser is said to be risk-seeking, and if it is
concave, the advertiser is said to be risk-averse. We show that
for risk-averse advertisers, pure per-click bidding is dominant, whereas pure
per-impression bidding is dominant when the advertisers are risk-seeking.

It is natural to assume that some advertisers may be either risk-averse or
risk-seeking. Risk-aversion models advertisers with tight budget constraints.
Risk-seeking advertisers also occur naturally in many settings; one example is
when advertisers are conducting experiments to identify high performance
advertising channels and keywords. Finding a high reward keyword may result in
a higher budget allocated to this keyword and more revenue in the future,
making the present utility function of winning this ad slot appear convex at
the present time.

We assume the advertisers are uninformed, which is equivalent to assuming the
advertiser and the auctioneer share a common prior, so that $\P_{j}=
\Q_{j}$. Essentially, the advertiser has no information and simply trusts the
auctioneer's prior\footnote{For the other extreme case of well-informed
  advertisers, there is no uncertainty, and hence the risk-averse and
  risk-seeking cases collapse to risk-neutral.}.  In this section, we focus on
a single advertiser, and drop the subscript corresponding to it.  Let $p =
\E[\P] = \E[\Q]$. As mentioned earlier, we assume that the auctioneer is
myopic as well ($\gamma_a = 0$), so that $q = \E[\Q] = p$.

Let $( m, c )$ denote the advertiser's bid, and let $v$
denote the true per-click valuation.  Let $I_R$ be the indicator
corresponding to whether the bidder gets the impression if $R_{-j} =
R$. The bidding strategy of the advertiser will attempt to maximize:

$$ I_R \cdot \max \left( \E \left[U \left(v \P - R \right) \right] ,
\E \left[ U \left(v \P - \frac{R \P }{p} \right) \right] \right) $$

In the above expression, the first term is the expected profit if the
impression is obtained based on the per-impression bid; and the latter
term is the expected profit if the impression is obtained based on the
per-click bid.  Our next lemma captures the structure of the dominant
strategy.
\begin{lemma}
\label{lem:risk}
If $U$ is concave, bidding $( 0, v )$ is a dominant
strategy. If $U$ is convex, the dominant strategy is of the form
$( m, 0 )$ for a suitably chosen $m$.
\end{lemma}
\begin{proof}
First consider the case when $v p < R$. In this case, regardless of
$U$, $\E \left[ U \left(v \P - \frac{R \P }{p} \right) \right] \le
0$. Therefore, to obtain positive profit, the bidder has to obtain the
impression based on his per-impression bid. In this case, the expected
profit is $ \E \left[U \left(v \P - R \right) \right] $. Note that
$\E[v \P - R] < 0$. Therefore if $g$ is concave:
$$ \E \left[U \left(v \P - R \right) \right] \le U \left(\E\left[v \P -
R \right] \right) \le U(0) = 0$$ Therefore, if $U$ is concave and $v
p < R$, then bidding $( 0, v )$ is a dominant
strategy. Since obtaining the impression based on the per-click bid
does not yield positive profit, if $U$ is convex, bidding $( m,
0 )$ with appropriately chosen $m$ is a dominant strategy.

The next and most interesting case is when $v p \ge R$.  Define random
variable $X = v \P - R$ and $Y = v \P - \frac{R \P }{p}$. First note that
$\E[X] = \E[Y] = v p - R \ge 0$. Further, the cumulative distribution
functions (CDFs) of $X$ and $Y$ cross exactly once, with the CDF of $X$ being
initially larger than the CDF of $Y$. This is sufficient to show that $Y$
Lorenz-dominates $X$. This implies that for $U$ being concave, $\E[U(Y)] \ge
\E[U(X)]$~\cite{mo:major79}, so that the advertiser only bids per
click. Further, if $U$ is convex, $\E[U(X)] \ge \E[U(Y)]$, so that the
advertiser only bids per impression.
\end{proof}

Our main result in this sub-section is the following property which gives a
single natural characterization of the optimum hybrid bid for both risk-averse
and risk-seeking advertisers. We will then show that for risk-seeking
advertisers ($U$ is strictly increasing and convex), the expected myopic
revenue of the auctioneer is larger in the {\sc Hybrid} auction compared to
the pure per-click auction, and for risk-averse advertisers, the {\sc Hybrid}
and per-click auctions coincide.
\begin{theorem}
Let $m^* = \max \{y | \E\left[ U\left(v \P - y\right) \right] \ge
0\}$. Bidding $( m^*,v)$ is a dominant strategy. Further,
the auctioneer's revenue in the {\sc Hybrid} scheme dominates the
revenue in the pure per-click scheme.
\end{theorem}
\begin{proof}
First consider the case when $U$ is concave. Then, $\E\left[ U\left(v \P
- p \right) \right] \le U\left(\E[v \P - v p] \right) =
0$. Therefore, $m^* \le v p$, so that bidding $( m^*,v)$
is equivalent to bidding $( 0, v)$, which is a dominant
strategy.

Next, when $U$ is convex, we have $m^* \ge v p$, so that bidding
$( m^*,v)$ is equivalent to bidding $( m^*,
0)$. The previous lemma shows that the dominant strategy is of
the form $( m,0)$. Since $m^*$ is the largest value of
$R_{-j}$ for which the advertiser makes a non-negative profit, bidding
$( m^*, 0)$ must be the dominant strategy.

The auctioneer's revenue in the {\sc Hybrid} scheme is the second
largest value of $\max(m^*_j, v_j p_j)$ while that in the per click
scheme is the second largest value of $v_j p_j$, which cannot be
larger.
\end{proof}

\section{Semi-Myopic Advertisers}
\label{sec:semi-myopic}
In this section, we remove the assumption that the advertisers are optimizing
some function of the profit at the current step. We now generalize to the case
where the advertisers optimize revenue over a time-horizon. We develop a
tractable model for the advertisers, and show a simple dominant strategy for
the advertisers, based on what we call the {\em bidding index}. Though this
strategy does not have a closed form in general, we show that in many natural
cases (detailed later) cases, it reduces to a natural pure per-click or pure
per-impression strategy that is socially optimal. Thus, our hybrid auctions
are flexible enough to allow the auctioneer and the advertiser to implement
complex dynamic programming strategies collaboratively, under a wide range of
scenarios. Neither per-impression nor per-click bidding can exhaustively mimic
the bidding index in these natural scenarios.

Recall that the true per-click value of advertiser $j$ is $v_j$, and that the
advertiser $j$ maintains a time-indexed distribution $\P_{jt}$ over the
possible values of the actual CTR $p$ that is updated whenever he receives an
impression. We assume advertiser $j$'s prior is updated based on the observed
clicks in a fashion similar to the auctioneer's prior, $\Q_{jt}$.

We assume the bidding strategy of the advertiser is {\bf semi-myopic},
which we define as follows: The advertiser has a discount factor
$\gamma_b$. The bid of an advertiser $j$ depends on its current state
$\langle v_j, \P_{jt}, \Q_{jt} \rangle$, and on $R_{-j}$ in a fashion
described next. At every step, the value of $R_{-j}$ is revealed. If
the advertiser $j$ got the impression the previous time step, the
value of $R_{-j}$ remains the same since the states of the other
advertisers remains the same, else it changes adversarially. The
optimization criterion of the advertiser is to maximize its discounted
expected gain (using discount factor $\gamma_b$) in the contiguous
time that it receives impressions (so that the value of $R_{-j}$
remains the same). We make the reasonable assumption that the
advertiser cannot optimize for a horizon beyond that, since the value
of $R_{-j}$ changes in an unknown fashion. Finally note that a myopic
advertiser is equivalent to assuming $\gamma_b = 0$.

\paragraph{Discussion.} The semi-myopic model is closely related to the MDP approach of analyzing repeated auctions; see for instance~\cite{athey,weber}. These works make the assumption that the {\em priors} $\P_j$ of the advertisers are public knowledge. However, this leads to somewhat perverse incentives in which the optimal strategy for an advertiser could be to underbid at the current time step in the hope that the priors of the other advertisers resolve to low values, and he then wins the auction on the remaining time steps at a lower price. However, note that if there are sufficiently many bidders, this scenario is unlikely to happen, and the bidder will attempt to win the auction at the current time slot. We make this explicit by making the following assumptions:
\begin{enumerate}
\item The bidder $j$ is aware of the revealed $R_{-j}$ values of the other bidders, but may not be aware of their prior distributions, which are usually private information.
\item The bidder only optimizes over the contiguous time horizon in which he receives the impressions. In this horizon, $R_{-j}$ is fixed, and further, this  removes the perverse incentive to under-bid described above.
\end{enumerate}

Note that in our model, the bidder is indeed {\em aware} of the current bids $R_{-j}$ of the other bidders. However, unlike the model in~\cite{athey,weber}, the optimization time-horizon of the bidder leads to the existence of a nicely specified dominant strategy. We hope that our modeling, that is only slightly more restrictive than ones considered in literature, but which have nice analytic properties, will be of independent interest in this and other contexts.

\subsection{The Dominant Bidding Index Strategy}
\label{sec:biddingindex}
We first show a bidding strategy that we term the {\em bidding index}
strategy, and show that it is weakly dominant in the class of {\em
semi-myopic} strategies.  The {\bf bidding index} $\B(v, \P, \Q)$ is
defined as follows: Suppose the advertiser's current prior is $\P$ and
the auctioneer's current prior is $\Q$. Denote the current time
instant as $t=0$.  Since the advertiser computes this index, we assume
the advertiser trusts his own prior but not the auctioneer's. For a
parameter $W$, define the following game between the advertiser and
the auctioneer with discount factor $\gamma_b$: At step $t \ge 0$,
suppose the advertiser has prior $\P_t$ (with mean $\E[\P_t] = p_t$)
and the auctioneer, $Q_{t}$ (with $q_t = f(\Q_t)$ being the
auctioneer's index), the advertiser can either stop the game, or
continue. If he continues, he gains $v p_t$ in expectation and pays
the auctioneer $W \min \left(1, \frac{p_t}{q_t}\right)$; the
difference is his gain. The advertiser's value for the game is the
expected discounted (according to $\gamma_b$) gain for the optimal
strategy.  Define $\W(v, \P, \Q)$ as the largest value of $W$ for
which the value of the game with initial priors $\P$ and $\Q$, is
positive. This value can easily be computed by dynamic programming,
much like the Gittins index. 

The {\bf bidding index} $\B(v, \P, \Q)$ is defined as:
$$ \B(v, \P, \Q) = \W(v,\P,\Q) \min \left(1,\frac{p_0}{q_0} \right)$$
This is the largest per impression price at time $t=0$ for which the
value of the above game is positive.

\paragraph{The Strategy:} At any time step, when the advertiser $j$'s
prior is $\P_{jt}$ with mean $p_{jt}$, and the auctioneer's prior is
$Q_{jt}$, with $q_{jt} = f(\Q_{jt})$, let $W_{jt} = \W(v_j, \P_{jt},
\Q_{jt})$ and $B_{jt} = \B(v_j, \P_{jt}, \Q_{jt})$. The {\bf bidding
index strategy} involves bidding $( B_{jt},
\frac{B_{jt}}{p_{jt}} )$.

It is clear that the bidding index strategy is well-defined for
$q_{jt}$ being an arbitrary function $f(\Q_{jt})$ chosen by the
auctioneer,and not just for $f$ being the Gittins index of $\Q_{jt}$
using discount factor $\gamma_a$. 

\begin{theorem}
\label{THM:BIDINDEX}
The bidding index strategy is (weakly) dominant in the class of
semi-myopic strategies.
\end{theorem}
\begin{proof}
Consider a sequence of time steps when advertiser $j$ gets the
impression; call this a phase. During this time, the value $R_{-j}$
used in the VCG scheme is fixed; denote this value $R^*$. Suppose at a
certain time step, the mean of the advertiser's prior is $p_{jt}$ and
the auctioneer computes $q_{jt}$. If the advertiser gets the
impression, the price he is charged in the VCG scheme is either $R^*$
per impression or $R^*/q_{jt}$ per click. The advertiser optimizes
this by paying $R^* \min(1, p_{jt}/q_{jt})$ in expectation per
impression. The state evolution is only conditioned on getting the
impression and not on the price paid for it.

Since the advertiser's strategy is semi-myopic, at any time step, the
bid should fetch him a non-negative expected profit for the rest of
the phase. This implies that $R^* \le W_{jt}$. There are two cases.

First, if $p_{jt} < q_{jt}$, the advertiser essentially bids $R_{jt} =
B_{jt} \frac{q_{jt}}{p_{jt}} = W_{jt} \ge R^*$, and receives the
impression at a per-click price of $\frac{R^*}{q_{jt}}$. Therefore,
the expected per impression price is $R^* \frac{p_{jt}}{q_{jt}} = R^*
\min \left(1,\frac{p_{jt}}{q_{jt}}\right)$.

Next, if $p_{jt} > q_{jt}$, the advertiser essentially bids $R_{jt} =
B_{jt} = W_{jt} \ge R^*$, and receives the impression at a per-impression
price of $R^* = R^* \min \left(1,\frac{p_{jt}}{q_{jt}}\right)$. 

Therefore, the bidding scheme ensures that the advertiser receives the
impression and makes the most possible profit in the rest of the
phase. Note finally that if $W_{jt} < R^*$, the maximum possible profit
in the rest of the phase is negative, and the bidding scheme ensures
the advertiser does not receive the impression.
\end{proof}

\subsection{Social Optimality of Bidding Index} 
\label{sec:indexsocial}
Suppose the global discount factor is $\gamma$. We define the socially
optimal strategy as follows: Suppose at time $t$, advertiser $j$ with
prior $\P_{jt}$ receives the impression resulting in value $v_j
p_{jt}$ for him. The socially optimal solution maximizes the infinite
horizon expected discounted value with discount factor $\gamma$. 

We show that the bidding index strategy implements the socially optimal
solution in each of the following two situations: (1) The advertiser and the
auctioneer share the same prior ($\P_{jt} = \Q_{jt}$), and either (1a) only
the advertisers are strategic ($\gamma_a = 0$ and $\gamma_b = \gamma$) or (1b)
only the auctioneer is strategic ($\gamma_a = \gamma$ and $\gamma_b = 0$); and
(2) The advertisers are certain about their CTR ($\P_j = p_j$) and (2a) the
auctioneer's index $q_{jt}$ is always at most $p_j$. The bidding index also
has a particularly simple form when the advertisers are certain, and (2b) the
auctioneer's $q_{j,t}$ is monotonically decreasing with $t$ and always larger
than $p_{j,t}$. In both (2a) and (2b), the bidding index strategy reduces to
bidding $( v_j p_j, v_j)$.

These scenarios are not arbitrarily chosen, and are the most illustrative
scenarios we could find. Scenario (1) corresponds to an advertiser and an
auctioneer that trust each other and hence have a common prior; in case (1a),
the auctioneer merely discloses its current estimate and trusts the
advertisers to bid in an optimal fashion, whereas in (1b) the advertisers
delegate the strategic decision making to the auctioneer. In scenario 2, the
advertisers have a definitive model of the CTR; in (2a), we model the case
where the auctioneer starts with an underestimate of the click-through rate
and hence the $q_{j,t}$ are always smaller than $p_{j}$ to which they will
hopefully converge as this advertisement is shown more times and the
auctioneer's prior gets sharpened, and in (2b) we model the mirror situation
where the $q_{j,t}$'s are always an over-estimate. It will be interesting to
find a general theorem about the bidding index that unifies all these diverse
scenarios.

In each of these cases, the bidding strategy can be implemented using either
per-impression or per-click bidding or both, but neither per-impression nor
per-click bidding can exhaustively mimic the bidding index in all scenarios.

\paragraph{Shared Priors.} When the advertisers are uncertain and
simply share the auctioneer's prior, we have $\P_{jt} = \Q_{jt}$. Let
$G_{jt}$ denote the Gittins index of $\P_{jt}$ with discount factor
$\gamma$. The socially optimal solution always gives the impression to
the advertiser with highest $v_j G_{jt}$ at time $t$. 

\begin{theorem}
  For shared priors, the bidding index strategy implements the socially
  optimal solution in the following two cases:
\begin{enumerate}
\item The advertisers are strategic, i.e., $\gamma_b = \gamma$, and the
  auctioneer is myopic, i.e., $\gamma_a = 0$.
\item The advertisers are myopic, i.e., $\gamma_b = 0$, and the auctioneer is
  strategic, i.e., $\gamma_a = \gamma$.
\end{enumerate}
\end{theorem}
\begin{proof}
  For the first part, we have $q_{jt} = p_{jt}$ since $\gamma_a =
  0$. Therefore, $\min\left(1,\frac{p_{jt}}{q_{jt}}\right) = 1$, so that the
  value $W_{jt}$ is the largest charge per impression so that the advertiser's
  discounted revenue is non-negative. This is precisely the definition of the
  Gittins index with discount factor $\gamma_b = \gamma$. Therefore, the
  bidding index strategy involves bidding $( v_j G_{jt}, v_j
  \frac{G_{jt}}{p_{jt}} )$. This can easily be seen to be equivalent to
  bidding either $( v_j G_{jt},0 )$ or $( 0, v_j
  \frac{G_{jt}}{p_{jt}} )$, and hence can be mimiced with either
  pure-impression or pure-click bidding. Also, we have $R_{jt} = v_j G_{jt}$,
  so that the bidding index implements the socially optimal strategy.

  For the second part, since the advertiser is myopic, the bidding index
  reduces to bidding $( v_j p_{jt}, v_j )$. Since $\gamma_a =
  \gamma$, we have $q_{jt} = G_{jt} \ge p_{jt}$. Therefore, $R_{jt} = v_j
  G_{jt}$, so that the bidding index implements the socially optimal solution;
  this can also be mimiced using per-click bidding but not per-impression
  bidding.
\end{proof}

\paragraph{Well -Informed Advertisers.} We next consider the case where the
advertisers are certain about their CTR $p_j$, so that $\P_{jt} = p_j$. The
socially optimal solution always gives the impression to the advertiser with
the largest $v_j p_j$. We show the following theorem:

\begin{theorem}
When $\P_{jt} = p_j$, then the bidding index strategy reduces to
bidding $( v_j p_j, v_j )$ in the following two scenarios:
\begin{enumerate}
\item The auctioneer's $q_{jt}$ is always at most $p_j$. In this case,
  the strategy is equivalent to bidding $( v_j p_j, 0 )$
  and is socially optimal.
\item The auctioneer's $q_{jt}$ is at least $p_j$, and is
  monotonically decreasing with $t$. 
\end{enumerate}
\end{theorem}
\begin{proof}
  When $p_j > q_{jt}$ for all $t$, we have
  $\min\left(1,\frac{p_j}{q_{jt}}\right) = 1$, so that the value $W_{jt}$ is
  the largest per-impression price for which the advertiser's discounted
  revenue is non-negative. This is precisely $v_j p_j$, so that the bidding
  index strategy reduces to bidding $( v_j p_j, v_j )$. This is
  clearly socially optimal. Since $q_{jt} < p_j$, this is equivalent to
  bidding $( v_j p_j, 0$, but can not be simulated using per-click bids.

  When $p_j \le q_{jt}$ for all $t$ and $q_{jt}$ is monotonically decreasing
  with $t$, the expected price $W_{jt} \frac{p_j}{q_{jt}}$ charged to the
  advertiser increases with time. At any time $t$, the advertiser maximizes
  $W_{jt}$ by setting it to $v_j q_{jt}$ and stopping after the first
  step. Therefore, $W_{jt} = v_j q_{jt}$, and $B_{jt} = v_j q_{jt}
  \frac{p_j}{q_{jt}} = v_j p_j$. Therefore, the bidding index strategy reduces
  to bidding $( v_j p_j, v_j )$; this is also equivalent to
  $( 0, v_j)$, but can not be simulated using per-impression bids.
\end{proof}

\newcommand{\m}{\bar{p}}

\section{Exploration by Advertisers}
\label{sec:exploration}
We now show a simple bidding strategy for a certain (i.e. well-informed)
advertiser to make the auctioneer's prior converge to the true CTR, while
incurring no extra cost for the advertiser; per-click bidding would have
resulted in the advertiser incurring a large cost.  More concretely, this
models the scenario where the advertiser has side information about the
advertisement's CTR but the auctioneer does not have a good prior, for
example, because the keyword may be obscure. The advertiser has incentive to
help the auctioneer ``learn'' the true CTR because this improves the
advertiser's chance of winning an ad slot in a pure per-click scheme.

To motivate why this is important, imagine a situation where the advertiser
would not get the slot if the scheme were pure per-click, and he were to bid
truthfully per-click, letting the auctioneer use his own estimate $q_j$ of the
CTR. Therefore, in the pure per-click scheme, the advertiser has to overbid on
the per-click valuation to get the slot enough number of times to make the CTR
used by the auctioneer converge to the true value; we show that this results
in loss in revenue for the advertiser. However, allowing for per impression
bids preserves truthfulness, and furthermore, helps the auctioneer ``learn''
the true CTR, while incurring no revenue loss to the advertiser.  This is our
final argument in support of hybrid auctions, and may be the most convincing
from an advertiser's point of view.

Formally, we  consider an advertiser that is certain about its CTR $p_j$, where
$v_j p_j > R_{-j}$ so that the advertiser can (and would like to) win the
auction but where  $q_{jt} < p_j$, and where the goal of the advertiser is to
make the auctioneer's prior converge to the true CTR. We show that the
advertiser can achieve this goal without any loss in revenue, whereas
achieving the same objective using per-click bidding would have resulted in a
large revenue-loss.  We assume the
auctioneer's prior is a Beta distribution.

We show a candidate strategy for an advertiser to make the Gittins index of
the auctioneer's distribution, $\Q_{jt} =$ Beta$(\alpha_{jt},\beta_{jt})$
converge close to its true CTR $p_j$ while incurring no loss in revenue. The
loss is defined as the value earned from actual clicks minus the amount paid
to the auctioneer.

We focus on a single advertiser and drop its subscript. For any
$\epsilon > 0$, suppose the advertiser's strategy is as follows:
During an ``explore'' phase, submit a bid of $( v p', v)$
where $p' = p(1-\epsilon)$, and then switch to bidding $( 0, v
)$.  During the explore phase, suppose the advertiser gets $T$
impressions on a price per impression basis resulting in $n$
clicks. Then the worst-case loss in revenue of the advertiser during
the explore phase is $v(T p' - n)$.  The ``explore'' phase stops when
the auctioneer's posterior mean of the distribution
Beta$(\alpha+n,\beta+T-n)$ is at least $p(1-\epsilon)$. Note that this
also implies that the Gittins index for Beta$(\alpha+n,\beta+T-n)$ is
at least $p(1-\epsilon)$ irrespective of the discount factor $\gamma$;
this in turn implies that by switching to pure per-click bidding, the
advertiser is assured that $q \ge p(1-\epsilon)$, so that bidding
$( 0,v )$ yields $R_j \ge v p (1-\epsilon)$.

\begin{claim}
  Suppose the advertiser knows its true CTR is $p$, and the auctioneer's
  initial prior is Beta$(\alpha,\beta)$. For any $\epsilon > 0$, the explore
  phase incurs no loss in revenue for the advertiser.
\end{claim}
\begin{proof}
  Let $T$ denote the random stopping time of the explore phase and suppose it
  results in $N$ clicks. First note that if $T > 0$, then $p(1-\epsilon) >
  \frac{\alpha}{\alpha+\beta}$. The posterior mean on stopping is $
  \frac{\alpha + N}{\alpha+\beta +T} \ge p(1-\epsilon)$, which implies $N/T >
  p (1-\epsilon)$. Therefore, $T p (1-\epsilon) - N < 0$, which shows there is
  no loss in revenue (provided $T$ is finite with probability 1, which follows
  from the law of large numbers in this case).
\end{proof}

Suppose $R_{-j} = v_j p_j (1-\epsilon)$. In a pure per-click bidding scheme,
the advertiser would have to bid at least $v_j (1-\epsilon)p_j/q_{jt}$ at time
$t < T$ with an expected loss (i.e. profit $-$ cost) of
$p_jv_j((1-\epsilon)p_j/q_{jt} - 1)$. For a myopic auctioneer with initial
prior $(1,\beta)$, the total loss of revenue for the advertiser till time $T$
is $\Omega(v_jp_j\beta)$ which can be arbitrarily large.

\section{Multi-Slot Auction}
\label{sec:multi}

In this section, we generalize the hybrid auction to multiple slots under the
standard separable CTR assumption, such that the resulting generalization is
truthful in a myopic setting analogous to Section~\ref{sec:myopic}. Assume
there are $K$ slots, where slot $i$ is associated with a CTR multiplier
$\theta_i \in [0,1]$. Slot $1$ is the topmost slot; since the CTRs decrease
with slot number, we have $1 = \theta_1 \ge \theta_2 \ge \cdots \ge \theta_K
\ge 0$. We will also define $\theta_{K+1} = 0$.

As before, advertiser $j$ and the auctioneer maintain priors on the CTR value
for this advertiser in ad slot 1. As before, we denote these priors as $\P_j$
and $\Q_j$ respectively. Let $p_j = \E[\P_j]$ be the expected CTR estimated by
the advertiser, and let $q_j = f(\Q_j)$ denote the Gittins index (or for that
matter, any other function) of the auctioneer's prior. Let $v_j$ denote the
true per-click valuation of advertiser $j$. Note that the priors $\P_j$ and
$\Q_j$ correspond to the estimated CTR for advertiser $j$ in the {\em first ad
  slot}, so that the expected CTR for the $i^{th}$ slot based on the
advertiser's estimate is $\theta_i p_j$.

Advertiser $j$ bids $( m_j, c_j )$, which is interpreted as the
per-impression and per-click bids for obtaining the {\em first slot}. The
auction is modeled after the laddered auction in~\cite{agm:auction06}, which
is equivalent to VCG under the separability assumption~\cite{edel-05}.  First,
compute the effective bid $R_{j} = \max\{m_j,c_jq_j\}$ for every advertiser as
  described in section~\ref{sec:hybrid}. Assume without loss of generality
  that there are $K+1$ advertisers, and that $R_1 \ge R_2 \ge \ldots \ge
  R_{K+1}$. Then, the auction proceeds as follows:
\begin{enumerate}
\item Advertiser $j$ is placed in slot $j$, for $1\le j \le K$. 
\item An ``effective charge'', $e_j$ is computed for advertiser $j$ as
  $e_j = \sum_{i=j}^{K} \left(\frac{\theta_i - \theta_{i+1}}{\theta_j}\right)
  R_{i+1}$.
\item If $m_j > c_jq_j$ then the advertiser is charged $e_j$ per impression;
  else it is charged $e_j/q_j$ per click.
\end{enumerate}

It is easy to see that $e_j\theta_j = R_j (\theta_j - \theta_{j+1}) + e_{j+1}
\theta_{j+1}$. Informally, advertiser $j$'s effective charge is the same as
the effective bid of the $(j+1)$-th advertiser for the additional click-rate
at the $j$-th position, and the same as the effective charge of the $(j+1)$-th
advertiser for the click-rate that would have already been realized at the
$(j+1)$-th position.

\begin{theorem}
  If $p_j = \E[\P_j]$ and the advertiser is myopic and risk-neutral, then
  regardless of the choice of $q_{j}$, the (strongly) dominant strategy is to
  bid $( v_j p_j, v_j )$.
\end{theorem}
The proof of the above theorem is obtained by extending the proof of
theorem~\ref{thm:truth} exactly along the line of the proof of truthfulness of
the laddered auction in~\cite{agm:auction06}, and is omitted from this
version.  This proof can also be obtained using the analysis of VCG with
probabilistic allocations, due to Myerson~\cite{myerson}.

\section{Conclusion}
Advertising is a major source of revenue for search engines and other
web-sites, and a major driver of innovation in web technology and services.
Advertising spots are typically sold on the web using auctions, and these
auctions have typically been either Cost-Per-Click (CPC), Cost-Per-Impression
(CPM), or Cost-Per-Action (CPA). We defined a single-slot hybrid auction,
which allows advertisers to enter per-impression as well as per-click bids. We
showed that this auction is truthful for risk-neutral, myopic advertisers, the
setting under which such auctions have typically been analyzed. When
advertisers are risk-seeking, or non-myopic, or when the advertiser has much
better information about the Click-Through-Rate (CTR) than the auctioneer, we
show that the hybrid auction offers stronger revenue guarantees and advertiser
flexibility than either pure CPC or CPM. The hybrid auction generalizes
naturally to multi-slot scenarios and is equally applicable to (CPM,CPA) or
(CPC,CPA) bidding. Finally, the hybrid auction is fully backwards compatible
with a CPC auction, in the sense that advertisers entering (optional)
per-impression bids in addition to per-click bids can seamlessly co-exist with
advertisers entering only per-click bids in the same auction.

\bibliographystyle{plain}

\begin{thebibliography}{10}

\bibitem{agm:auction06}
G.~Aggarwal, A.~Goel, and R.~Motwani.
\newblock Truthful auctions for pricing search keywords.
\newblock {\em Proceedings of the seventh ACM conference on Electronic
  Commerce}, pages 1--7, June 2006.

\bibitem{athey}
S.~Athey and I.~Segal.
\newblock An efficient dynamic mechanism.
\newblock 2007.
\newblock Available at {\tt http://kuznets.fas.harvard.edu/\~{}athey}.

\bibitem{weber}
A.~Bapna and T.~Weber.
\newblock Efficient dynamic allocation with uncertain valuation.
\newblock {\em Working Paper, Department of Management Science and Engineering,
  Stanford University}, 2005.

\bibitem{edel-05}
B.~Edelman, M.~Ostrovsky, and M.~Schwarz.
\newblock Internet advertising and the generalized second price auction:
  Selling billions of dollars worth of keywords.
\newblock {\em American Economic Review}, 97(1):242--259, 2007.

\bibitem{g:gittins79}
J.~C. Gittins.
\newblock Bandit processes and dynamic allocation indices.
\newblock {\em J Royal Statistical Societe Series B}, 14:148--167, 1979.

\bibitem{lpsv:auctions07}
S.~Lahaie, D.~Pennock, A.~Saberi, and R.~Vohra.
\newblock Sponsored search auctions.
\newblock {\em In Algorithmic Game Theory, edited by Nisan, Roughgarden,
  Tardos, and Vazirani}, 2007.

\bibitem{mo:major79}
A.W. Marshall and I.~Olkin.
\newblock {\em Inequalities: theory of majorization and its applications}.
\newblock Academic Press (Volume 143 of Mathematics in Science and
  Engineering), 1979.

\bibitem{msvv-05}
A.~Mehta, A.~Saberi, U.~Vazirani, , and V.~Vazirani.
\newblock Adwords and generalized online matching.
\newblock {\em Proceedings of the 46th {IEEE} Symposium on Foundations of
  Computer Science}, 2005.

\bibitem{myerson}
R.~B. Myerson.
\newblock Optimal auction design.
\newblock {\em Mathematics of Operations Research}, 6(1):58--73, 1981.

\bibitem{rdr:ctr07}
M.~Richardson, E.~Dominowska, and R.~Ragno.
\newblock Predicting clicks: estimating the click-through rate for new ads.
\newblock {\em Proceedings of the 16th international conference on World Wide
  Web}, pages 521--530, 2007.

\bibitem{v:google06}
H.~Varian.
\newblock Position auctions.
\newblock {\em International Journal of Industrial Organization}, October 2006.

\bibitem{w:gittins92}
R.~Weber.
\newblock On the gittins index for multiarmed bandits.
\newblock {\em Annals of applied probability}, 2(4):1024--1033, 1992.

\end{thebibliography}

\end{document}